\documentclass[aps,prd,preprintnumbers,nofootinbib]{revtex4}
\usepackage[utf8]{inputenc}
\usepackage{empheq}
\usepackage{bm}
\usepackage[normalem]{ulem}
\usepackage{latexsym}
\usepackage{color}
\usepackage{dcolumn}
\usepackage{amsmath,amsfonts,amssymb}
\usepackage{graphicx,epsfig}
\usepackage{psfrag}
\usepackage{amsthm}
\usepackage[pdftex]{hyperref}
\usepackage{amsmath}
\usepackage[toc,title,page]{appendix}
\usepackage{natbib}
\usepackage{amsfonts}
\def\be{\begin{eqnarray}}
\def\ee{\end{eqnarray}}
\def\tred{\textcolor{red}}

\begin{document}

\title{The Rise of Cosmological Complexity: Saturation of Growth and Chaos}
\author{Arpan Bhattacharyya} \email{abhattacharyya@iitgn.ac.in}
\affiliation{
Indian Institute of Technology,
Gandhinagar,Gujarat 382355, India}
\author{Saurya Das}\email{saurya.das@uleth.ca}
\affiliation{Theoretical Physics Group, \\
Department of Physics and Astronomy,\\
University of Lethbridge, 4401 University Drive,\\
Lethbridge, Alberta T1K 3M4, Canada}
\author{S. Shajidul Haque}\email{shajid.haque@uct.ac.za}
\affiliation{High Energy Physics, Cosmology \& Astrophysics Theory Group \\and The Laboratory for Quantum Gravity \& Strings,\\
Department of Mathematics and Applied Mathematics,\\
University of Cape Town, South Africa} 
\author{Bret Underwood}\email{bret.underwood@plu.edu}
\affiliation{Department of Physics,\\
Pacific Lutheran University,\\
Tacoma, WA 98447}

\date{\today}

\vspace{-2cm}
\begin{abstract}
We compute the circuit complexity of scalar curvature perturbations on FLRW cosmological backgrounds with fixed equation of state $w$ using the language of squeezed vacuum states.
Backgrounds that are accelerating and expanding, or decelerating and contracting, exhibit features consistent with chaotic behavior, including linearly growing complexity.
Remarkably, we uncover a bound on the growth of complexity for both expanding and contracting backgrounds $\lambda \leq \sqrt{2} \ |H|$, similar to other bounds proposed independently in the literature.
The bound is saturated for expanding backgrounds with an equation of state more negative than $w = -5/3$, and for contracting backgrounds with an equation of state larger than $w = 1$.
For expanding backgrounds that preserve the null energy condition, de Sitter space has the largest rate of growth of complexity (identified as the Lyapunov exponent), and we find a scrambling time that is similar to other estimates up to order one factors.
\end{abstract} 

\maketitle
\section{Introduction}

The relatively new concept called {\it circuit complexity} -- conceptually defined as the minimum number of quantum gates necessary to construct a desired target state from a given reference state -- has found several interesting applications in recent years.
For example, it may play an important role in probing physics behind the horizon of an eternal anti-de Sitter (AdS) black hole through the ``complexity $=$ volume'' and ``complexity $=$ action'' proposals \cite{com1,com2,com3,com4}.
Recent studies have also explored different methods of calculating circuit complexity based on the position space wavefunction or covariance matrix \cite{MyersCC,Guo:2018kzl,Khan:2018rzm,Hackl:2018ptj,Bhattacharyya:2018bbv,Camargo:2018eof,Ali:2018aon}, using the geometric technique pioneered by Nielsen \cite{NL3}, and have extended the idea of complexity to quantum field theory \cite{MyersCC,Chapman:2017rqy,Caputa:2017yrh,Bhattacharyya:2018wym,Caputa:2018kdj,Bhattacharyya:2019kvj,Caputa:2020mgb,Flory:2020eot,Erdmenger:2020sup}.
More generally, complexity can serve as one of several diagnostics for determining if a quantum system displays quantum chaos \cite{Balasubramanian:2019wgd,me3,Yang:2019iav}.

We are interested here in the application of circuit complexity to cosmology, in order to gain new insights about cosmological evolution and structure formation from a quantum information theoretic perspective.
In particular, cosmological perturbations in Friedmann-Lema\^{i}tre-Robertson-Walker (FLRW) backgrounds can be described in the language of two-mode squeezed states \cite{Grishchuk,Albrecht,Martin1,Martin2}.
The squeezed state formalism allows the straightforward use of the geometric techniques described above for computing the cosmological complexity of cosmological perturbations, first described in \cite{us}.
The preliminary study \cite{us} only studied two simple expanding cosmological backgrounds: de Sitter (or cosmological constant-dominated), and radiation-dominated.
For expanding de Sitter backgrounds, the cosmological complexity is small at early times while the mode is within the horizon, but grows linearly with the logarithm of the scale factor after the mode exits the horizon.
Because the linear growth of complexity is a characteristic \cite{me3} of quantum chaos, \cite{us} proposed that cosmological perturbations in de Sitter space are (quantum) chaotic. From this perspective the Lyapunov exponent, given by the slope of the growing complexity, is set by the de Sitter Hubble constant, and the scrambling time is set by the horizon exit time of the mode of interest.
The behavior of complexity for expanding radiation solutions found in \cite{us}, however, is quite the opposite. While the mode is outside the horizon, the complexity is decreasing linearly with the logarithm of the scale factor, and ``freezes-in'' once the mode re-enters the horizon at late times.
This suggests that cosmological perturbations in an expanding radiation background are not chaotic, in contrast with the de Sitter case.

In this work, we aim to extend this analysis of the cosmological complexity of cosmological perturbations to a broader set of backgrounds.
In particular, we will analyze the behavior of the squeezing parameters and complexity for backgrounds with arbitrary constant equation of state $w$, encompassing both accelerating and decelerating backgrounds.
We find that the squeezing and complexity for accelerating and decelerating (expanding) backgrounds behave qualitatively similar to their de Sitter and radiation counterparts from \cite{us}: accelerating solutions show signs of quantum chaos, with complexity growing as modes exit the horizon, while the complexity for decelerating solutions decays until the mode re-enters the horizon.
Interestingly, we find that the slope of the growing complexity for accelerating backgrounds increases with decreasing equation of state $w$ until reaching a ``threshold'' value $w = -5/3$, after which the slope is fixed. 
This implies a bound on the rate of change of complexity, which we calculate to be $d{\mathcal C}/dt \leq \sqrt{2}\ H$. 
This is an independent derivation of a similar bound on the growth rate of complexity \cite{Maldacena:2015waa}, and implies a fundamental limit on the Lyapunov exponent of a quantum chaos system.
For backgrounds that satisfy the null energy condition, de Sitter has the largest complexity growth, and therefore the largest Lyapunov exponent.
We also extract a scrambling time for cosmological perturbations, finding rough agreement with other results in the literature \cite{Dvali:2013eja,gary}.

Further, we will extend our analysis to include contracting as well as expanding backgrounds, finding that the squeezing and complexity for decelerating contracting backgrounds behave qualitatively similar to accelerating contracting backgrounds.
Contracting backgrounds will obey an identical bound on the growth rate of complexity, saturating the bound for equations of state larger than $w = 1$.

In Section \ref{sec:SqueezedCosmo} we review some important details about the description of cosmological perturbations as two-mode squeezed states, and find approximate analytic solutions for the squeezing parameters for expanding and contracting backgrounds. 
We confirm these analytic solutions by qualitatively comparing them to numerical solutions to the squeezing equations of motion for a variety of equations of state.
In Section \ref{sec:Complexity} we use the squeezing solutions found in Section \ref{sec:SqueezedCosmo} to find approximate analytic solutions for the cosmological complexity, uncovering the bound on complexity growth and the scrambling time.
We conclude with some discussion, speculation, and potential future work in Section \ref{sec:Discussion}.

\section{Squeezed Cosmological Perturbations}
\label{sec:SqueezedCosmo}

We begin our analysis by reviewing the description of cosmological perturbations \cite{Mukhanov} as two-mode squeezed states, following \cite{us} (see also \cite{Grishchuk,Albrecht,Martin1,Martin2}).
We take as our background a spatially flat Friedmann-Lemaitre-Robertson-Walker (FLRW) metric
\be
ds^2 = -dt^2 + a(t)^2 d\vec{x}^2 = a(\eta)^2 \left(-d\eta^2+d\vec{x}^2\right)\, ,
\label{FLRW}
\ee
where we have introduced the conformal time $\eta$.
For simplicity we will consider our matter content to consist of a fundamental scalar field $\varphi$ with a canonical kinetic term and potential\footnote{It would be interesting to study the effect that more general matter content has on the behavior of squeezed states and complexity, as in \cite{Albrecht}, though we leave these investigations to future work.} $V(\varphi)$. On the background (\ref{FLRW}) the scalar field is homogeneous and time-dependent $\varphi_0(t)$; the scalar field potential and kinetic energy source the evolution of the scalar field $a(\eta)$.

On this background we will consider fluctuations of a scalar field $\varphi(x) = \varphi_0(t) + \delta\varphi(t,x)$ with the metric
\be
ds^2 = a(\eta)^2 \left[-(1+2\psi(x,\eta))d\eta^2 + (1-2\psi(x,\eta)) d\vec{x}^2\right]\, .
\ee
In terms of the curvature perturbation ${\mathcal R} = \psi+\frac{H}{\dot \varphi_0} \delta \varphi$, where a dot denotes a derivative with respect to cosmic time $t$, and the Hubble parameter $H = \dot a/a$, the action takes the following simple form \cite{Mukhanov}
\be
S = \frac{1}{2}\int dt\, d^3x\, a^3 \frac{\dot \phi^2}{H^2} \left[\dot {\mathcal R}^2 - \frac{1}{a^2} \left(\partial_i {\mathcal R}\right)^2\right]\, .
\ee
By using the Mukhanov variable $v \equiv z {\mathcal R}$
where $z \equiv a\, \sqrt{2\epsilon}$, with $\epsilon = -\dot H/H^2 = 1-{\mathcal H}'/{\mathcal H}^2$, the action can be transformed into the following form:
\be
S = \frac{1}{2} \int d\eta\, d^3x \left[v'^2 - (\partial_i v)^2 + \left(\frac{z'}{z}\right)^2 v^2 - 2 \frac{z'}{z} v' v\right]\, ,
\label{CosmoAction}
\ee
where the prime denotes a derivative with respect to conformal time and ${\mathcal H} = a'/a$.
The last term in (\ref{CosmoAction}) is usually removed by an integration by parts, but we will retain this form for our analysis.
This action represents perturbations of a scalar field coupled to an external time-varying source\footnote{A virtually identical-looking expression can also be derived for tensor perturbations with the replacement $z'/z \rightarrow a'/a$, and our results will hold for these types of perturbations as well.}.

We 
define the usual creation and annihilation operators in terms of the Fourier modes of the perturbation
\be
\hat v_{\vec{k}} = \frac{1}{\sqrt{2k}} \left(\hat c_{\vec{k}} + \hat c^\dagger_{-\vec{k}}\right), \hspace{.2in}
\hat v_{\vec{k}}' = -i\frac{k}{2} \left(\hat c_{\vec{k}} - \hat c_{-\vec{k}}^\dagger\right)\, .
\label{CosmoCreationOperators}
\ee
The unitary evolution ${\mathcal U_{\vec{k}}}$ of a state can be parameterized in the factorized form
\cite{Grishchuk,Albrecht}
\be
\hat {\mathcal U}_{\vec{k}} = \hat{\mathcal S}_{\vec{k}}(r_k,\phi_k) \hat{\mathcal R}_{\vec{k}}(\theta_k)\, ,
\ee
where $\hat{\mathcal R}_{\vec{k}}$ is the two-mode rotation operator
\be
\hat{\mathcal R}_{\vec{k}}(\theta_k) \equiv {\rm exp}\ \left[-i\theta_k(\eta) (\hat c_{\vec{k}} \hat c_{\vec{k}}^\dagger + \hat c_{-\vec{k}}^\dagger \hat c_{-\vec{k}})\right]\, ,
\ee
written in terms of the rotation angle parameter $\theta_k(\eta)$ and $\hat{\mathcal S}_{\vec{k}}$ is the two-mode squeeze operator
\be
\hat {\mathcal S}_{\vec{k}}(r_k,\phi_k) \equiv {\rm exp}\ \left[\frac{r_k(\eta)}{2} \left(e^{-2i\phi_k(\eta)} \hat c_{\vec{k}} \hat c_{-\vec{k}} - e^{2i\phi_k(\eta)} \hat c_{-\vec{k}}^\dagger \hat c_{\vec{k}}^\dagger\right)\right]\, ,
\ee
written in terms of the squeezing parameter $r_k(\eta)$ and squeezing angle $\phi_k(\eta)$.
For our case, the rotation operator and rotation angle $\theta_k$ are not important, therefore, will be dropped in the subsequent analysis.
Also, since the squeezing equations of motion will only depend on the magnitude $k$ of the wavenumber $\vec{k}$, we have suppressed the vector notation on the subscripts of these parameters.


The two-mode vacuum is annihilated by $\hat c_{\vec{k}}$
\be
\hat c_{\vec{k}} |0\rangle_{\vec{k},-\vec{k}} = 0, \hspace{.2in} \forall\ \vec{k}\, .
\ee
The two-mode squeeze operator acting on this vacuum then results in a two-mode \emph{squeezed} vacuum state
\begin{equation}
    |\Psi_{sq}\rangle_{\vec{k},-\vec{k}} = \hat {\mathcal S}(r_k,\phi_k)_{\vec{k}} |0\rangle_{\vec{k}} = \frac{1}{\cosh r_k} \sum_{n=0}^{\infty} (-1)^n e^{-2 i n \phi_k} \tanh^n r_k\, |n_{\vec{k}}; n_{-\vec{k}}\rangle\, ,
    \label{psi1}
\end{equation}
where the two-mode excited state is
\be
|n_{\vec{k}}; n_{-\vec{k}}\rangle = \sum_{n=0}^\infty \frac{1}{n!} \left(\hat c_{\vec{k}}^\dagger\right)^n \left(\hat c_{-\vec{k}}^\dagger\right)^n\, |0\rangle_{\vec{k},-\vec{k}}\, .
\ee
The time evolution of the squeezing parameters $r_k(\eta),\phi_k(\eta)$ is determined by the Schr\"odinger equation
and leads to the differential equations for $r_k(a)$ and $\phi_k(a)$
with the scale factor $a$ as the independent variable
\begin{eqnarray}
\frac{dr_k}{da} &=& -\frac{1}{a}\,\cos(2\phi_k) \, ;
\label{eom1} \\
\frac{d\phi_k}{da} &=& \frac{k}{a{\cal H}}
+ \frac{1}{a}\, \coth(2r_k) \sin(2\phi_k)~.
\label{eom2}
\end{eqnarray}
Note that ${\cal H}(a) \equiv a'/a$ is implicitly defined in terms of the scale factor $a$. 
In the following subsections we will specify the background $a(\eta)$ for a fixed equation of state for expanding and contracting cosmologies, which will allow us to solve the equations of motion (\ref{eom1})-(\ref{eom2}) as a function of the scale factor.

\subsection{Expanding Backgrounds}

We will begin by considering expanding FLRW backgrounds (\ref{FLRW}) with a fixed equation of state $p = w\rho$. The scale factor as a function of conformal time depends on whether the background is accelerating or decelerating through the convenient parameter $\beta \equiv -\frac{2}{1+3w}$
\begin{equation}
    a(\eta) = \left(\frac{\eta_0}{\eta}\right)^\beta = 
    \begin{cases}
    \left(\frac{\eta_0}{\eta}\right)^\beta, & -\infty < \eta < 0, ~ \eta_0 < 0 ~ \mbox{for accelerating backgrounds $\beta > 0$ ($w < -1/3$)} \\
    \left(\frac{\eta}{\eta_0}\right)^{|\beta|}, & 0 < \eta < \infty, ~ \eta_0 > 0 ~ \mbox{for decelerating backgrounds $\beta < 0$ ($w > -1/3$)}\, .
    \end{cases}
    \label{ExpandingScaleFactor}
\end{equation}
Note that the parameter $\beta = -2/(1+3w)$ is positive for accelerating solutions and negative for decelerating solutions.
We will treat the marginal $w=-1/3$ case separately at the end of this subsection.
The equations of motion (\ref{eom1})-(\ref{eom2}) can now be written entirely in terms of the scale factor by inverting (\ref{ExpandingScaleFactor}), giving
\begin{eqnarray}
\frac{dr_k}{da} &=& -\frac{1}{a}\,\cos(2\phi_k) \, ;
\label{eom3} \\
\frac{d\phi_k}{da} &=& \frac{k|\eta_0|}{|\beta|}\,
\frac{1}{a^{1+1/\beta}}
+ \frac{1}{a}\, \coth(2r_k) \sin(2\phi_k)~.
\label{eom4}
\end{eqnarray}
The behavior of the solutions (\ref{eom3})-(\ref{eom4}) depend crucially on whether the background is accelerating or decelerating, so we will study these cases separately.


Beginning with accelerating backgrounds $w < -1/3$, we will see in our analysis below that the behavior of the solutions depends qualitatively on the numerical value of $\beta$, which specifies the equation of state of the background.
An accelerating background equation of state that obeys the {\bf null energy condition} $-1 \leq w < -1/3$ has the range $1 \leq \beta < \infty$, where $\beta = 1$ corresponds to a de Sitter background $w = -1$.
Equations of state that {\bf violate} the null energy condition $ w < -1$ thus fall in the range $0 < \beta < 1$.

At sufficiently early times $a \ll 1$ perturbation modes start inside the horizon for all accelerating solutions.
Solutions to (\ref{eom3})-(\ref{eom4}) start with small squeezing $r_k \ll 1$ and approximately constant squeezing angle
\be
r_k(a) &\approx& \frac{\beta}{2k|\eta_0|} a^{1/\beta}\, ; \label{AccelSmallScaleSolnSqueeze}\\
\phi_k(a) &\approx& -\frac{\pi}{4} + \frac{1}{4k|\eta_0|} a^{1/\beta}\, ,
\label{AccelSmallScaleSolnAngle}
\ee
up to subleading corrections of higher order in $a$.
The mode exits the horizon when $k \sim {\mathcal H}$ or equivalently
when $k|\eta_0| \sim \beta a^{1/\beta}$, which is when the first
term in the squeezing angle equation of motion (\ref{eom4}) balances
the coefficient of the second term.

After the mode exits the horizon the squeezing angle asymptotically approaches $-\pi/2$ as $a \gg 1$ and the squeezing parameter grows logarithmically with the scale factor and becomes large $r_k \gg 1$.
The form of the solutions turns out to depend on the value of $\beta$
\be
r_k(a) &\approx & \ln a\, ; \label{AccelSqueezeSoln} \\
\phi_k(a) &\approx& -\frac{\pi}{2} \ + \ \underbrace{
\frac{k|\eta_0|}{2\beta-1}\, \frac{1}{a^{1/\beta}}}_{\text{Inhomogeneous}} +  \underbrace{\frac{B}{a^2}}_{\text{Homogeneous}},
\label{AccelAngleSoln}
\ee
where $B$ is an arbitrary constant dependent on initial conditions, and we have included both the homogeneous (independent of $k$) and homogeneous (dependent on $k$) subleading solutions for the squeezing angle.
For $\beta > 1/2$, corresponding to $w > -5/3$, the inhomogeneous solution for the squeezing angle dominates for $a \gg 1$. Alternatively, for $\beta < 1/2$, corresponding to $w < -5/3$, the homogeneous solution dominates the subleading behavior of the squeezing angle.
For certain initial conditions these two terms can
switch roles temporarily -- see Appendix \ref{app:IntermediateBehavior}.
Oddly, this transition between the behaviors does not occur at the boundary of null energy violation $w = -1$, corresponding to $\beta = 1$.
The transition at $\beta = 1/2$ can be traced to the factor of two in the $\sin(2\phi_k)$ term of (\ref{AccelAngleSoln}), corresponding to the two-mode structure of the squeezed state, though its deeper physical origin is less clear.
For $\beta = 1/2$, we need to modify the solution (\ref{AccelAngleSoln}) to include an additional factor of $\ln a$ since the powers of the homogeneous and inhomogeneous solutions are the same.
Putting all of these conditions together, we have the solutions
\be
r_k(a)
= \begin{cases}
\ln(a) + 
{\cal O} \left( \frac{1}{a^{2/\beta}}\right) & \beta \geq 1/2 \\
\ln(a) + 
{\cal O} \left( \frac{1}{a^{4}}\right) & \beta < 1/2,
\end{cases}\, ,
\label{AccelLargeScaleSolnSqueeze}
\ee
for the squeezing parameter and
\be
\phi_k(a)
= \begin{cases}
-\frac{\pi}{2} + 
\frac{k|\eta_0|}{2\beta-1}\, \frac{1}{a^{1/\beta}} & \beta > 1/2\ (w > -5/3)\\
-\frac{\pi}{2} + \frac{B}{a^2} + \frac{k|\eta_0|}{\beta}\frac{\ln a}{a^2} & \beta = 1/2\ (w = -5/3)\\
- \frac{\pi}{2} + \frac{B}{a^2} & \beta < 1/2\ (w < -5/3)
\end{cases}\, ,
\label{AccelLargeScaleSolnAngle}
\ee
for the squeezing angle.

\begin{figure}[t]
\includegraphics[width=.45 \textwidth]{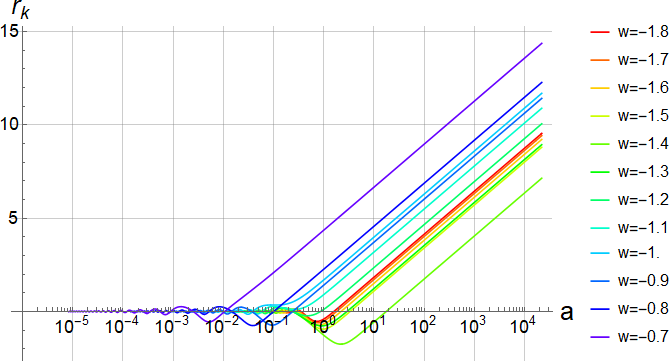} \hspace{.2in}
\includegraphics[width=.45\textwidth]{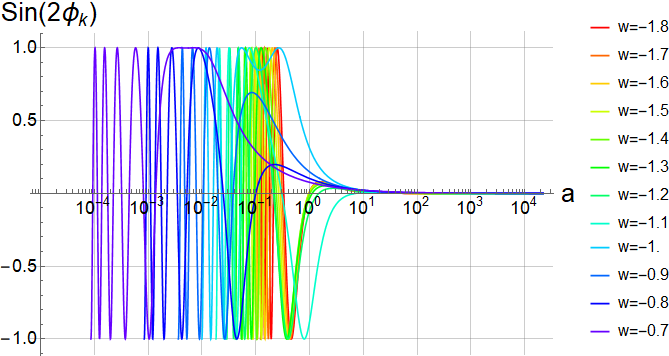}
\caption{(Left) The squeezing parameter $r_k$ for expanding accelerating solutions grows at late times as $\ln a$ in a universal way, so that all equations of state have the same slope in the growth of the squeezing parameter. (Right) The squeezing angle, shown as $\sin(2\phi_k)$, for expanding accelerating solutions approaches $\phi_k \rightarrow -\pi/2 + {\mathcal O}(1/a^n)$ at late times, for some power $n$ determined in (\ref{AccelLargeScaleSolnAngle}), so that $\sin(2\phi_k)$ approaches zero at late times. As discussed in the text, the power $n$ depends on the equation of state $w$, and determines the slope of the complexity growth, as discussed in Section \ref{sec:Complexity}.}
\label{fig:AccelExpandingSqueeze}
\end{figure}

We can see the general behavior of the early-time solutions (\ref{AccelSmallScaleSolnSqueeze})-(\ref{AccelSmallScaleSolnAngle}) and late-time solutions
(\ref{AccelLargeScaleSolnSqueeze})-(\ref{AccelLargeScaleSolnAngle})
in the numerical solutions to the equations of motion (\ref{eom3})-(\ref{eom4}) in Figure \ref{fig:AccelExpandingSqueeze}, shown for different choices
of the equation of state $w$ (equivalently different values of $\beta$).
In particular we note that at early times the squeezing parameter $r_k(a)$ is small, then grows as $\sim \ln a$ after the mode exits the horizon, with the same slope for all equations of state, while the squeezing angle $\phi_k$ after the mode exits the horizon asymptotically approaches $-\pi/2$ in a way that depends on the equation of state.
This behavior is qualitatively similar to that of de Sitter solutions ($w=-1$) found in \cite{Albrecht,us}.

Before we consider the decelerating solutions we point out that for an equation of state above the ``threshold'' value $w = -5/3$ identified in (\ref{AccelLargeScaleSolnSqueeze})-(\ref{AccelLargeScaleSolnAngle}) the subleading behavior of the squeezing angle $\phi_k(a)$ depends on the equation of state through the power of the scale factor $a^{1/\beta}$.
As the equation of state decreases past $w = -5/3$ the subleading behavior of the squeezing angle $\phi_k(a)$ changes according to (\ref{AccelLargeScaleSolnAngle}), so that the power of the subleading behavior of $\phi_k(a)$ no longer depends on the equation of state.
As we will see in Section \ref{sec:Complexity}, the subleading behavior of the squeezing angle will determine the slope of the growth of complexity in the superhorizon limit. The slope of the growth of the complexity, then, will depend on the equation of state until it reaches the threshold value $w=-5/3$, after which it becomes independent of the equation of state.

%
%
%
%

Perturbations for decelerating solutions $w > -1/3$ ($\beta < 0$) begin outside the horizon for sufficiently early times $a \ll 1$.
In general, the squeezing parameter grows when a mode is superhorizon, so the squeezing is growing at early times for decelerating solutions to (\ref{eom3})-(\ref{eom4})
\be
r_k(a) &\approx& r_0 + \ln \left(a/a_0\right)\, ; \label{DecelLargeScaleSolnSqueeze} \\
\phi_k(a) &\approx& -\frac{\pi}{2} + \frac{k\ \eta_0}{|\beta|(|\beta|+2)} a^{1/|\beta|}\, ,
\label{DecelLargeScaleSolnAngle}
\ee
where we introduced a fiducial scale factor $a_0$ where the squeezing parameter takes the value $r_k(a_0) = r_0$.
Notice that the squeezing solutions here again grow universally as $\sim \ln a$ for all equations of state. On the other hand, the subleading term in the solution for the squeezing angle depends on the equation of state through the coefficients and power of the scale factor depending on $|\beta|$. Unlike for accelerating backgrounds, however, there is no ``threshold'' value of the equation of state beyond which the qualitative nature of the solution for the squeezing angle changes. Instead, the solutions (\ref{DecelLargeScaleSolnSqueeze})-(\ref{DecelLargeScaleSolnAngle}) apply to all equations of state.
\begin{figure}[t]
\includegraphics[width=.45 \textwidth]{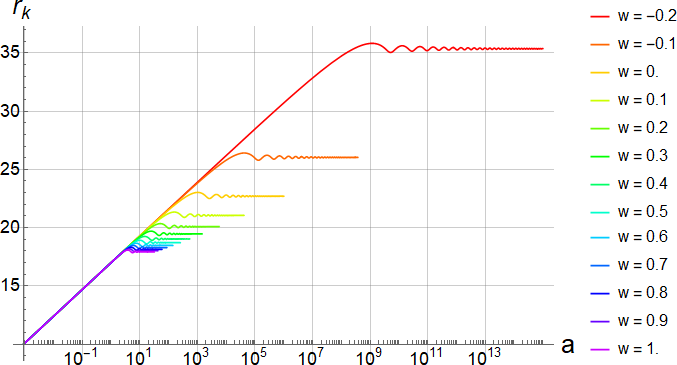} \hspace{.2in}
\includegraphics[width=.45\textwidth]{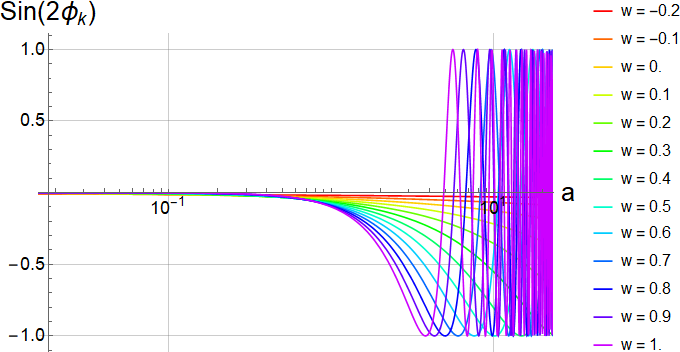}
\caption{(Left) The squeezing parameter $r_k$ for expanding decelerating solutions grows at early times (when the mode is outside the horizon) as $\ln a$ in a universal way, so that all equations of state have the same slope in the growth of the squeezing parameter. The squeezing then ``freezes in'' when the mode crosses the horizon, which depends on the equation of state. (Right) The squeezing angle, shown as $\sin(2\phi_k)$, for expanding decelerating solutions is approximately $\phi_k \rightarrow -\pi/2 + {\mathcal O}(a^n)$ at early times, for some power $n$, so that $\sin(2\phi_k)\approx$ zero when the mode is superhorizon. Once the mode crosses the horizon $\phi_k$ begins to grow, leading to oscillations in $\sin(2\phi_k)$. As discussed in the text, the power $n$ depends on the equation of state $w$, and determines the slope of the complexity decay, as discussed in Section \ref{sec:Complexity}.}
\label{fig:ExpandNonAccelerSqueezing}
\end{figure}

At late times, after the mode re-enters the horizon
the squeezing ``freezes-in'' and the squeezing angle begins to grow, so that solutions to (\ref{eom3})-(\ref{eom4}) in this regime take the form
\be
r_k(a) &\approx& r_*\, ; \label{DecelSmallScaleSolnSqueeze} \\
\phi_k(a) &\approx& k\ \eta_0\ a^{1/|\beta|}\, , \label{DecelSmallScaleSolnAngle}
\ee
for some fixed value of the squeezing $r_*$, determined by the value of the squeezing at horizon re-entry.

Together, this behavior for early and late times is qualitatively similar to the solution for squeezing in a radiation-dominated expanding background discussed in \cite{us}.
In particular, in Figure \ref{fig:ExpandNonAccelerSqueezing} we see that the numerical solutions a wide range of equations of state of interest follow the qualitative behavior described above: squeezing grows initially as the mode begins outside of the horizon, then ``freezes-in'' after the mode re-enters the horizon. The squeezing angle begins approximately constant for early times, then evolves after the mode re-enters the horizon.

Finally, let's consider the marginal $w=-1/3$ case, which is neither accelerating nor decelerating.
The scale factor in conformal time takes the form
\be
a(\eta) = a_0\, e^{\eta/\eta_0},~ \eta_0 > 0, ~ -\infty < \eta < \infty \, .
\ee
The conformal Hubble parameter then becomes constant ${\cal H} = 1/\eta_0$, so the equations of motion (\ref{eom1})-(\ref{eom2}) become
\be
\frac{dr_k}{da} &=& -\frac{1}{a} \cos(2\phi_k)\, ; \label{eom5} \\
\frac{d\phi_k}{da} &=& \frac{k\ \eta_0}{a} + \frac{1}{a} \coth(2 r_k) \sin(2\phi_k)\, . \label{eom6}
\ee
Unlike the accelerating and decelerating backgrounds, however, a co-moving mode for $w=-1/3$ stays fixed relative to the horizon and does not exit or re-enter the horizon at early or late times.
The nature of the solutions to (\ref{eom5})-(\ref{eom6}) are then controlled by the quantity $k\ \eta_0$. For $k\  \eta_0 \gg 1$, the mode is inside the horizon, and the squeezing remains small and the squeezing angle is approximately fixed.
These solutions will not be interesting to us because we are interested in solutions in which the squeezing is large and growing, which we will see correspond to growing complexity.
For $k\ \eta_0 \ll 1$, the mode is outside the horizon, and the equations of motion (\ref{eom5})-(\ref{eom6}) for the squeezing variables have the solution for $a \gg 1$
\be
r_k(a) &\approx& \ln a\, ; \label{ExpandingLargeScaleMarginalSqueeze} \\
\phi_k(a) &\approx& -\frac{\pi}{2} + \frac{1}{2} k\ \eta_0 + \frac{B}{a^2}\, , \label{ExpandingLargeScaleMarginalAngle}
\ee
where again $B$ is a constant set by the initial conditions.
As with previous superhorizon solutions, we see again that the squeezing parameter grows while the squeezing angle asymptotically approaches a constant, for $a \gg 1$. Note, however, that the constant in this limit is shifted slightly from $-\pi/2$ for $w=-1/3$ by an amount $\sim k\ \eta_0 \ll 1$, which will have implications for the late-time behavior of the complexity.

\subsection{Contracting Backgrounds}

Now we will consider contracting cosmological backgrounds with fixed equation of state.
The scale factor for a contracting universe with fixed equation of state can be written as
\begin{equation}
    a(\eta) = \left(\frac{\eta_0}{\eta}\right)^\beta = 
    \begin{cases}
    \left(\frac{\eta_0}{\eta}\right)^\beta, & 0 < \eta < \infty, ~ \eta_0 > 0 ~ \mbox{for accelerating backgrounds $\beta > 0$ ($w < -1/3$)} \\
    \left(\frac{\eta}{\eta_0}\right)^{|\beta|}, & -\infty < \eta < 0, ~ \eta_0 < 0 ~ \mbox{for decelerating backgrounds $\beta < 0$ ($w > -1/3$)}\, ,
    \end{cases}
    \label{ContractingScaleFactor}
\end{equation}
where $\beta =-2/(1+3w)$ as before.
Again, we will treat the marginal $w=-1/3$ case separately.
The equations of motion (\ref{eom1})-(\ref{eom2}) can now be written as
\begin{eqnarray}
&& \frac{dr_k}{d\eta} = \frac{\beta}{\eta}\,\cos(2\phi_k)\, ; \label{reometa} \\
&& \frac{d\phi_k}{d\eta} = k - \frac{\beta}{\eta}\,\coth(2r_k)\,\sin(2\phi_k)~,
\label{phieometa}
\end{eqnarray}
in terms of the conformal time $\eta$, or rewritten as
\begin{eqnarray}
&& \frac{dr_k}{da} = -\frac{1}{a}\,\cos(2\phi_k)~; \label{eom7} \\
&& \frac{d\phi_k}{da} = -\frac{k\,|\eta_0|}{|\beta|}\frac{1}{a^{1+1/\beta}} + \frac{1}{a}\,\coth(2r_k)\,\sin(2\phi_k) ~, \label{eom8}
\end{eqnarray}
in terms of the scale factor $a$. Note that ${\cal H} < 0$, which changes the sign of the first term on the right-hand side of (\ref{eom8}).
Since we are in a contracting universe, the scale factor is large $a \gg 1$ at early times and becomes small $a \ll 1$ for late times.
In order to avoid confusion, we will present solutions below in terms of both $\eta$ and $a$, keeping in mind that $\eta$ always increases, while $a$ is decreasing.

\begin{figure}[t]
\includegraphics[width=.45\textwidth]{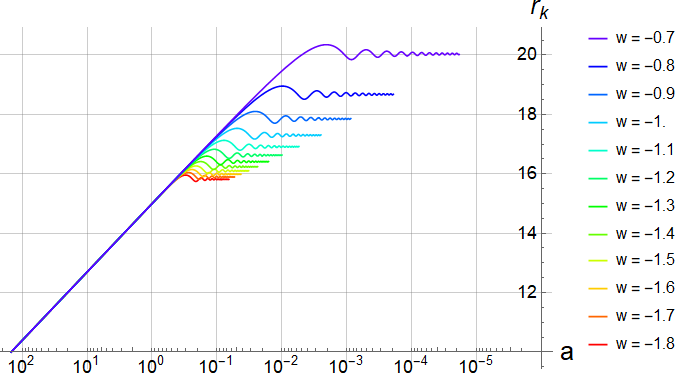}\hspace{.2in}\includegraphics[width=.45\textwidth]{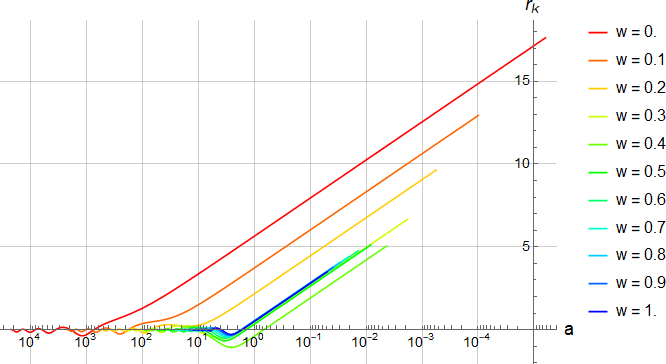}
\caption{(Left) The squeezing parameter $r_k$ for an accelerating contracting universe grows until the mode exits the horizon, after which it ``freezes in''. Note that since the universe is contracting, the scale factor evolves from large values to small values, thus the flow of time is to the right. (Right) The squeezing parameter $r_k$ for a decelerating contracting universe is approximately zero until the mode exits the horizon, after which the squeezing grows as $\sim \ln a$, universally for all equations of state $w$.}
\label{fig:ContractingSqueezing}
\end{figure}

Since a contracting universe is a time-reversal of an expanding universe, we expect most of the same qualitative features of the solutions from the previous section to be present in these solution as well.
However, while in an expanding accelerating background modes begin inside the horizon and exit the horizon as the universe evolves, the opposite is true for a contracting universe. Thus, we expect the squeezing solutions for contracting accelerating backgrounds to resemble those of the expanding decelerating backgrounds.
In particular, early-time $a \gg 1$ ($\eta/\eta_0 \ll 1$) solutions for accelerating backgrounds have large squeezing $r_k \gg 1$ and take the form
\be
r_k &\approx & r_0 - \beta \ln \left(\frac{\eta}{\eta_0}\right) = r_0 - \ln a\, ; \label{ContractAccelLargeScaleSqueeze} \\
\phi_k &\approx & \left(\frac{k}{1+2\beta}\right) \eta = \left(\frac{k\, \eta_0}{1+2\beta}\right) a^{-1/\beta} \, . \label{ContractAccelLargeScaleAngle}
\ee
%
After the mode re-enters the horizon as the universe contracts the squeezing parameter ``freezes-in'', as it did for the expanding decelerating solutions
\be
r_k &\approx & r_*\, ; \label{ContractAccelSmallScaleSqueeze} \\
\phi_k &\approx& k\, \eta  = k\,\eta_0\, a^{-1/\beta}\, , \label{ContractAccelSmallScaleAngle}
\ee
where $r_*$ is the value of the squeezing parameter at horizon crossing.
The numerical solutions of (\ref{eom7})-(\ref{eom8}) shown in the left panel of Figure \ref{fig:ContractingSqueezing} for several accelerating equations of state $w < -1/3$ also illustrate this ``freeze-in'' behavior for the squeezing parameter $r_k$ (note that the horizontal axis starts at large values of the scale factor on the left, as is appropriate for a contracting universe).

For a contracting decelerating universe, a typical mode begins inside the horizon at early times $a \gg 1$ ($\eta \rightarrow -\infty$) and exits the horizon as the universe evolves.
At early times, then, the solution for the squeezing variables will resemble that of the expanding accelerating universe
\be
r_k &\approx& 
\frac{|\beta|}{2\, k\, \eta} = \frac{|\beta|}{2\, k\, \eta_0} a^{-1/|\beta|}\, ; \label{ContractDecelSmallScaleSqueeze} \\
\phi_k &\approx& \frac{\pi}{4} + \frac{1}{4\, k\, \eta} = \frac{\pi}{4} + \frac{1}{4\, k\, \eta_0} a^{-1/|\beta|}\, , \label{ContractDecelSmallScaleAngle}
\ee
with small squeezing $r_k \ll 1$.
Similarly, at late times $a \ll 1$ the mode is outside of the horizon, and the squeezing is large and growing $r_k \gg 1$.
As with the expanding universe, the dominant subleading behavior of the squeezing angle changes as the equation of state (equivalently the parameter $\beta$) crosses a ``threshold'' value.
As a result, we have the solutions
\be
r_k &\approx & -|\beta| \ln\left(\frac{\eta}{\eta_0}\right) = -\ln a\, ; \label{ContractDecelLargeScaleSqueeze} \\
\phi_k &\approx &
\left\{\begin{array}{l l r l l}
\frac{k}{1-2|\beta|} \eta &= \frac{k\, |\eta_0|}{1-2|\beta|} a^{1/|\beta|} & & \beta < -1/2\ &(-1/3 < w < 1)\, \\
B\ \eta + k\ \eta\ \ln(\eta/\eta_0) &= B\ |\eta_0|\ a^{1/|\beta|} - \frac{k\ |\eta_0|}{|\beta|}\ a^{1/|\beta|}\ \ln a & &\beta = -1/2\ &(w=1)\, \\
B\ \eta^{2|\beta|} &= B\ a^2  &-1/2 < &\beta < 0 \ &(w > 1)
\end{array}\right. \, .\label{ContractDecelLargeScaleAngle}
\ee
Again, the numerical solutions of (\ref{eom7})-(\ref{eom8}) shown in the right panel of Figure \ref{fig:ContractingSqueezing} for several decelerating equations of state agree with these solutions for the squeezing parameter $r_k$, particularly at late times where the squeezing grows as $\sim (-\ln a)$ as $a$ decreases (again note that the horizontal axis has a decreasing scale factor as one moves to the right since we have a contracting background).

Finally, we will consider the marginal $w=-1/3$ background for completeness.
%
The scale factor in conformal time takes the form
\be
a(\eta) = a_0\, e^{\eta/\eta_0} = a_0\, e^{-\eta/|\eta_0|},~ \eta_0 < 0, ~ -\infty < \eta < \infty \, .
\ee
Just like the expanding case, the conformal Hubble parameter then becomes constant ${\cal H} =- 1/|\eta_0|$, therefore, the equations of motion (\ref{eom1})-(\ref{eom2}) become
\be
\frac{dr_k}{da} &=& -\frac{1}{a} \cos(2\phi_k)\, ; \label{eom5v2} \\
\frac{d\phi_k}{da} &=& - \frac{k |\eta_0|}{a} + \frac{1}{a} \coth(2 r_k) \sin(2\phi_k)\, , \label{eom6v2}
\ee
and a co-moving mode for $w=-1/3$ stays fixed relative to the horizon and does not exit or re-enter the horizon at early or late times.
Once again, for $k\, |\eta_0| \gg 1$, the mode is inside the horizon, and the squeezing parameter is small and the squeezing angle is fixed $\phi_k \approx -\pi/4$. For $k\, |\eta_0| \ll 1$, the mode is outside the horizon, and the equations of motion (\ref{eom5v2})-(\ref{eom6v2}) for the squeezing variables have the solution at late times $a \ll 1$
\be
r_k(a) &\approx& -\ln a\, ; \label{ContractLargeScaleMarginalSqueeze} \\
\phi_k(a) &\approx& \frac{1}{2} k |\eta_0| + B\ a^2\, , \label{ContractLargeScaleMarginalAngle}
\ee
where again $B$ is a constant set by the initial conditions.
As the universe contracts the squeezing is large and growing, and the squeezing angle approaches a small, non-zero constant.

\section{Complexity for the Squeezed State}
\label{sec:Complexity}

In the previous section, we saw that it is natural to describe the evolution of scalar cosmological perturbations as a two-mode squeezed vacuum state. 
Now we will investigate the complexity of the squeezed cosmological perturbations. For our quantum circuit, a natural reference state is the two-mode vacuum state $|0\rangle_{\vec{k},-\vec{k}}$, while our target state will be the squeezed two-mode vacuum state $|\Psi_{sq}\rangle_{\vec{k},-\vec{k}}$ in (\ref{psi1}).
As in \cite{MyersCC}, we will express the reference and target states as Gaussian wavefunctions. In that direction we first define a set of auxiliary ``position'' and ``momentum'' variables
\be
\hat q_{\vec{k}} \equiv  \frac{1}{\sqrt{2k}} \left(\hat c_{\vec{k}}^\dagger + \hat c_{\vec{k}}\right), \hspace{.2in}
\hat p_{\vec{k}} \equiv  i\sqrt{\frac{k}{2}}\left(\hat c_{\vec{k}}^\dagger - \hat c_{\vec{k}}\right)\, ,
\ee
which are conjugate variables $[\hat q_{\vec{k}},\hat p_{\vec{k}'}] = i \delta^3(\vec{k}-\vec{k}')$.
Notice that the main difference between the ``position'' $\hat q_{\vec{k}}$ and the Fourier mode $\hat v_{\vec{k}}$ given in (\ref{CosmoCreationOperators}) is that the former is defined with respect to a raising operator of $\vec{k}$ instead of $-\vec{k}$.

The two-mode vacuum state's wavefunction, defined by $\hat c_{\vec{k}} |0\rangle_{\vec{k},-\vec{k}} = 0$, has the usual gaussian form
\be
\psi_R(q_{\vec{k}},q_{-\vec{k}})= \langle q_{\vec{k}},q_{-\vec{k}} | 0\rangle_{\vec{k},-\vec{k}} = \left(\frac{k}{\pi}\right)^{1/4}\, e^{-\frac{k}{2} (q_{\vec{k}}^2 + q_{-\vec{k}}^2)}\, .
\label{reference}
\ee

To calculate the wavefunction corresponding to the squeezed state (\ref{psi1}) we note that the following combination annihilates $|\Psi_{sq}\rangle_{\vec{k},-\vec{k}}$
\be
\left(\cosh r_k\ \hat c_{\vec{k}} + e^{-2i\phi_k} \sinh r_k\ \hat c_{-\vec{k}}^\dagger\right) |\Psi_{sq}\rangle_{\vec{k},-\vec{k}} = 0\, .
\ee
Using this we can calculate the ``position-space'' form of the wavefunction \cite{Martin2}
\begin{equation} \label{state1}
\Psi_{sq} (q_{\vec{k}}, q_{-\vec{k}})= \langle q_{\vec{k}},q_{-\vec{k}}|\Psi_{sq}\rangle_{\vec{k}} =  \frac{e^{A(q_{\vec{k}}^2+q_{-\vec{k}}^2)-B q_{\vec{k}} q_{-\vec{k}}}}{\cosh r_k \sqrt{\pi} \sqrt{ 1- e^{-4 i \phi_k} \tanh^2 r_k}}\, ,
\end{equation}
where the coefficients $A$ and $B$ are functions of the squeezing parameter $r_k$ and squeezing angle $\phi_k$
\begin{equation}
    A= \frac{k}{2} \left( \frac{e^{-4 i \phi_k} \tanh^2 r_k +1}{e^{-4 i \phi_k} \tanh^2 r_k -1} \right)\, ,\hspace{.2in} B= 2k \left( \frac{e^{-2 i \phi_k} \tanh r_k }{e^{-4 i \phi_k} \tanh^2 r_k -1}\right)\, .
    \label{ABSqueezed}
\end{equation}
For computing complexity we will use the  Nielsen's geometric method \cite{NL3} and follow the approach of \cite{MyersCC}, which we will simply call the {\it circuit complexity}. For our quantum circuit we will use the vacuum as the reference state (\ref{reference}) and two mode squeezed state as the target state (\ref{state1}). By using the ``geodesic'' weighting we obtain the following expression for complexity ${\mathcal C}$
\begin{eqnarray}
    {\mathcal C}(k) &=&\frac{1}{2} \sqrt{ \left(\ln \left|\frac{\Omega_{\vec{k}}}{\omega_{\vec{k}}}\right| \right)^2+ \left(\ln \left|\frac{\Omega_{-\vec{k}}}{\omega_{-\vec{k}}}\right| \right)^2+ \left(\tan^{-1} \frac{\text{Im}\ \Omega_{\vec{k}}}{\text{Re}\ \Omega_{\vec{k}}}\right)^2+ \left(\tan^{-1} \frac{\text{Im}\ \Omega_{-\vec{k}}}{\text{Re}\ \Omega_{-\vec{k}}}\right)^2},
\label{complexity2}
\end{eqnarray}
where $\Omega_{\vec{k}}=-2 A+B$, $\Omega_{-\vec{k}}=-2A-B$, and $\omega_{\vec{k}} = \omega_{-\vec{k}} = k/2$. The inverse tangent term in the above expression is necessary when the frequency is complex, see \cite{me1}, although we will find that it doesn't play an important role in the qualitative behavior of the complexity.
Using (\ref{ABSqueezed}) in (\ref{complexity2}) we can obtain relatively simple closed-form expressions for complexity for the general two-mode squeezed vacuum state relative to the un-squeezed vacuum \cite{us}
\begin{eqnarray}
{\mathcal C} &=& \frac{1}{\sqrt{2}}\,
\sqrt{\left| \ln \left|
\frac{1+ e^{-2i\phi_k} \,\tanh{r_k}}{1 - e^{-2i\phi_k} \,\tanh{r_k}}
\right| 
\right|^2
+ \left| \tan^{-1} \left(2\sin 2 \phi_k \sinh r_k \cosh r_k \right)\right|^2}\, .\label{ComplexitySqueeze}
\end{eqnarray}
Different choices of the weighting function lead to similar looking expressions as (\ref{ComplexitySqueeze}) with slightly different numerical factors. For example, when a {\it linear} weighting function is used the terms in (\ref{ComplexitySqueeze}) are not added in quadrature, but are simply added together, with a coefficient a factor of $\sqrt{2}$ larger \cite{us}.
Since several different weighting factors agree on the general form of the complexity, but not on the numerical factor out front, we will regard the overall multiplicative factor of the complexity (\ref{ComplexitySqueeze}) as uncertain.

In the rest of this section we will analyze the behavior of the complexity (\ref{ComplexitySqueeze}) as a function of the scale factor by inserting the solutions for the squeezing parameters $r_k(a),\phi_k(a)$ found in Section \ref{sec:SqueezedCosmo} for expanding and contracting backgrounds.

\subsection{Complexity for Expanding FLRW Backgrounds}

We start by analyzing (\ref{ComplexitySqueeze}) for expanding backgrounds.
As noted in Section \ref{sec:SqueezedCosmo}, the qualitative nature of the solutions for the squeezing parameters depends on whether the FLRW background is accelerating or not.

For accelerating FLRW backgrounds, at early times $a \ll 1$ the squeezing parameter is small $r_k \ll 1$ and the squeezing angle is approximately constant $\phi_k \approx -\pi/4$, as found in (\ref{AccelSmallScaleSolnSqueeze})-(\ref{AccelSmallScaleSolnAngle}).
In this limit the complexity (\ref{ComplexitySqueeze}) vanishes ${\mathcal C}\rightarrow 0$ for all accelerating equations of state.

\begin{figure}[t]
\includegraphics[width=.7\textwidth]{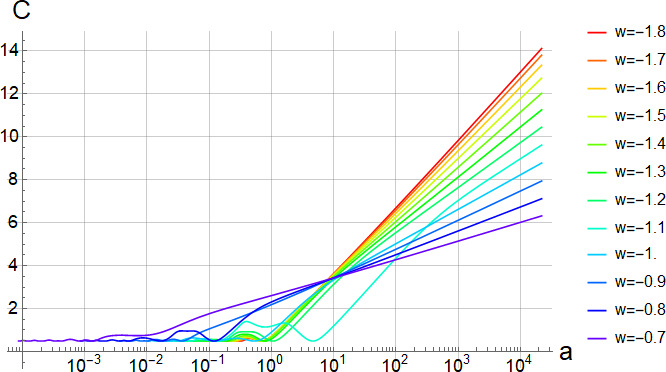}
\caption{The complexity for accelerating expanding solutions grows linearly with $\ln a$ at late times. The slope grows with decreasing equation of state until $w = -5/3$, after which the slope saturates, so that the growth of complexity is bounded by $d{\mathcal C}/dt \leq \sqrt{2} H$.}
\label{fig:AcceleratingExpandingComplexity}
\end{figure}

At late times $a \gg 1$, the squeezing parameter is now large and growing, and the squeezing angle has changed so that $\phi_k \approx -\pi/2$.
In this limit, the first term in (\ref{ComplexitySqueeze}) dominates over the second, so the complexity is approximately
\begin{eqnarray}
{\mathcal C} & \approx & \frac{1}{\sqrt{2}}\,
{\left| \ln \left|
\frac{1+ e^{-2i\phi_k} \,\tanh{r_k}}{1 - e^{-2i\phi_k} \,\tanh{r_k}}
\right| 
\right|} ~.\label{com1}
\end{eqnarray}
Since $r_k \gg 1$, $\tanh r_k \approx 1$ in (\ref{com1}). The dependence of (\ref{com1}) on the scale factor then primarily comes from the subleading behavior of the squeezing angle.
Inserting the solutions (\ref{AccelLargeScaleSolnSqueeze})-(\ref{AccelLargeScaleSolnAngle})
into (\ref{com1}), and recalling that $\beta = -2/(1+3w)$,
we find that the complexity grows as $\sim \ln a$ at late times for all accelerating backgrounds
\be
{\mathcal C}
\approx \begin{cases}
- \frac{1+3w}{2\sqrt{2}}\,\ln(a) + \mbox{const.} & -5/3 < w<-1/3 \\
\sqrt{2}\ \ln(a) - \sqrt{2}\ \ln(\ln(a))& w=-5/3 \\
\sqrt{2}\ \ln(a) + \mbox{const.} & w<-5/3
\end{cases}\, ,
\label{ExpandAccelComplexity}
\ee
where the double-log for $w=-5/3$ arises from the $\ln a$ in subleading behavior of the squeezing angle (\ref{AccelLargeScaleSolnAngle}).
Note that the extremely late time limit of the threshold $w=-5/3$ solution approaches $\sim \sqrt{2}\ \ln(a)$. However, for even fairly large values of the scale factor $a\sim 10^{10}$ the double-log term is still approximately $\sim 10\%$ the leading-order term, and contributes less than $1\%$ only for extremely large values of the scale factor $a > 10^{200}$.

Putting the early and late time behaviors together, then, we expect that for accelerating FLRW backgrounds the complexity for a cosmological perturbation will be approximately zero until the mode exits the horizon.
After the mode exits the horizon the complexity grows linearly with $\ln a$ (equivalently the number of e-folds), with a slope given by (\ref{ExpandAccelComplexity}).
As discussed in \cite{us}, the complexity for this mode will continue to grow indefinitely until the cosmological background itself changes and becomes non-accelerating.
Unlike other studies of complexity, then, we find that the complexity does not reach a maximum value in a finite amount of time, but instead grows without bound.
This appears to be the case for our system because the expanding background acts as a classical time-dependent source coupled to the cosmological perturbations, so that we don't have a closed system.
Figure \ref{fig:AcceleratingExpandingComplexity} shows numerical solutions of the complexity (\ref{ComplexitySqueeze}) for accelerating FLRW backgrounds with fixed equation of state, where we see clearly the general qualitative behavior for the complexity described above.
In \cite{me3}, it was argued that this behavior for the complexity -- a period of approximately constant complexity followed by linear growth -- is
a characteristic of \emph{quantum chaos}. 
Similar behavior for the complexity in an expanding de Sitter background was found already in \cite{us}, where it was argued that cosmological perturbations in a de Sitter background are chaotic. Here we extend this argument: 
based on the behavior of the complexity, cosmological perturbations in all
expanding, accelerating backgrounds display characteristic signs of quantum chaos.

Quantum chaos is often described by two parameters, the \emph{scrambling time} and the \emph{Lyapunov exponent}.
Viewed through this lens, the duration of the approximately constant part of complexity corresponds to the scrambling time scale, while the slope of the linear portion behaves as a Lyapunov exponent. 

For our cosmological perturbations, the complexity is constant (and approximately zero) until the modes exit the horizon; the horizon exit time thus sets the scale for the scrambling time.
Horizon exit occurs when $k \sim {\mathcal H}$, which corresponds to $k\ |\eta_0| \sim \beta\  a^{1/\beta}$. 
The number of e-folds up to scrambling for a particular $k$-mode then becomes $N_e^{scramb} \sim \beta \ln(k|\eta_0|/\beta)$.
The number of e-folds required for scrambling is maximized (for fixed $k\ |\eta_0|$) at $\beta = \ln(k|\eta_0|) - 1$.
It would be interesting to study this further to see if there are physical consequences to such a ``slow scrambler.''
For de Sitter space $w = -1$, we have $|\eta_0| = H_{dS}^{-1}$, where $H_{dS}$ is the de Sitter Hubble constant.
Taking the mode to have a Planck scale wavelength, the number of e-folds up to ``maximum scrambling'' (scrambling of all modes up to the Planck scale) are then bounded by
\be
N_e^{scramb} \leq \ln\left(\frac{M_{pl}}{H_{dS}}\right)\,.
\ee
Interestingly, this is of the same order (up to a factor of two) as the scrambling time for de Sitter space proposed in \cite{Dvali:2013eja}, which was also found in a study of the out-of-time-order correlator for de Sitter space in \cite{gary}.
It is also identical to the timescale of the recently proposed Trans-Planckian Censorship Conjecture (see \cite{Bedroya:2019snp,Bedroya:2019tba}, among others), since it is simply the timescale for a Planck-scale wavelength to exit the horizon.
Translating the scrambling condition into cosmic time, the scrambling time is approximately 
\be
t_{scramb} \sim \begin{cases}
t_0\ \left[\frac{k\ t_0\ |1+3w|}{2}\right]^{-3(1+w)/(1+3w)} & w \neq -1 \\
H_{dS}^{-1}\ \ln \left(\frac{k}{H_{dS}}\right) & w = -1
\end{cases}
\, ,
\label{ScramblingTime}
\ee
where $t_0$ is the typical age of the universe for equation of state $w$.
Again, we see that for de Sitter space our proposed scrambling time (\ref{ScramblingTime}) is of the same order (up to a factor of two) as the scrambling times proposed in \cite{Dvali:2013eja,gary}.
More generally, for $w\rightarrow -1/3$, the scrambling time (\ref{ScramblingTime}) vanishes $t_{scramb} \rightarrow 0$. Alternatively, for large negative equation of state $w \rightarrow -\infty$ the scrambling time (\ref{ScramblingTime}) goes to zero as $|w|^{-3}$.
The maximum scrambling time is therefore somewhere in between, and is set by $k\ |\eta_0|$.

\begin{figure}[t]
\includegraphics[width=.45 \textwidth]{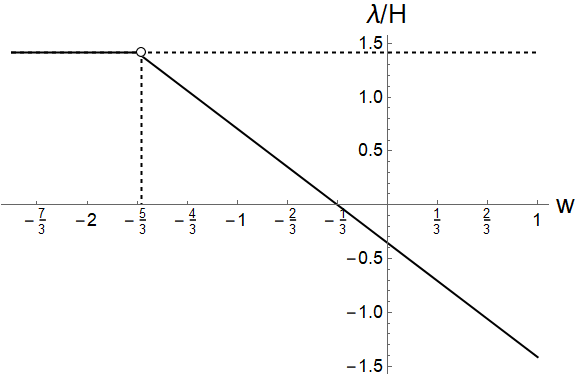} \hspace{.2in} \includegraphics[width=.45 \textwidth]{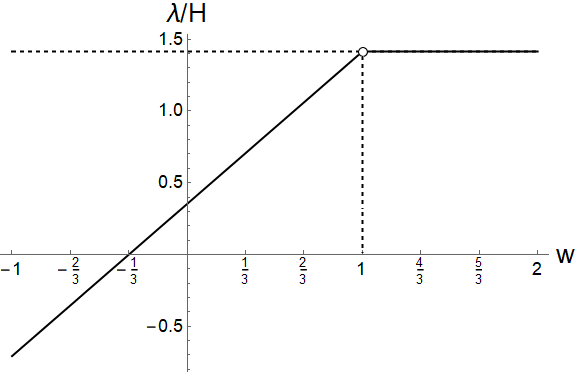}
\caption{The slope of complexity growth $\lambda = d{\mathcal C}/dt$, evaluated in terms of the Hubble parameter $H = da/dt$, for an expanding (left) and contracting (right) FRLW background with equation of state $w$.}
\label{saturation}
\end{figure}

Interestingly, the slope of the growth of complexity with $\ln a$ changes at the ``threshold'' value of the equation of state $w = -5/3$. 
The importance of this is perhaps better seen by constructing the slope $\lambda$ of the growth of the complexity with respect to cosmic time, 
which using (\ref{ExpandAccelComplexity}) becomes
\be
\frac{d {\mathcal C}}{dt} \equiv \lambda
\approx \begin{cases}
- \frac{1+3w}{2\sqrt{2}}\, H, & -5/3 < w < -1/3 \\
\sqrt{2}\ H\ \left(1-\frac{1}{\ln (a)}\right) & w = -5/3 \\
\sqrt{2}\, H, &  w <  -5/3.
\end{cases}\, ,
\label{lambdaExpanding}
\ee
where $H = \dot a/a$ is the Hubble parameter defined with respect to cosmic time.
For equations of state less negative than $w=-5/3$ (including de Sitter space $w=-1$), the slope of the complexity depends on the value of the equation of state, vanishing as we approach $w \rightarrow -1/3$, and increasing as the equation of state becomes more negative.
For equations of state more negative than $w=-5/3$, the slope {\bf saturates} to a fixed value $\lambda \sim \sqrt{2}\ H$ independent of the equation of state.
The behavior of the slope $\lambda$ for expanding backgrounds is shown in the left panel of Figure \ref{saturation}.
The open dot for $w=-5/3$ in Figure \ref{saturation} reflects the slow convergence of the slope (\ref{lambdaExpanding}) for this threshold equation of state due to the double-log behavior
found in (\ref{ExpandAccelComplexity}).
Notice that for backgrounds that obey the null energy condition, de Sitter has the largest slope and therefore the largest Lyapunov exponent $\lambda/H = 1/\sqrt{2}$, so de Sitter space appears to be maximally chaotic (in the sense of having the largest Lyapunov exponent) among null energy condition satisfying backgrounds.
A similar result for the Lyapunov exponent for de Sitter space in $(2+1)$-dimensions was found in the context of the out-of-time-order correlator \cite{gary}. Though it is unclear at the moment the relationship between the two very different approaches, it is intriguing that both give rise to very similar descriptions of quantum chaos in de Sitter space.

The behavior of the slope of the complexity
(\ref{lambdaExpanding}) implies that the slope is {\bf bounded}
\be
\frac{d {\mathcal C}}{dt} \equiv \lambda \leq \sqrt{2}\ H\, .
\label{lambdaExpandingBound}
\ee
The saturation of the slope of complexity growth is in agreement with general expectations \cite{Maldacena:2015waa} of the behavior of the growth of complexity. In particular, \cite{Maldacena:2015waa} conjectured that the growth of complexity is bounded from above by the temperature
\be
\lambda \leq 2\pi T \sim H\, ,
\label{lambdaBound}
\ee
where we used the temperature of an expanding background $T \sim H/2\pi$ \cite{Gibbons:1977mu}.
Up to an ${\mathcal O}(1)$ factor, we find an independent determination of this bound: the slope of the growth of complexity saturates at $w=-5/3$ to a value $\lambda \sim {\mathcal O}(1) H$.
Interestingly, this ${\mathcal O}(1)$ factor becomes precisely one, in exact agreement with (\ref{lambdaBound}), if we don't ``double-count'' the $+\vec{k}$ and $-\vec{k}$ modes in (\ref{complexity2}), or alternatively use a {\it linear} weighting functional instead of the geodesic weighting we have chosen here.
Since the precise weighting function and formulation of complexity for quantum field theories is still in development, we will just note this minor difference in passing.
Regardless of the precise numerical factor, it is fascinating that our analysis independently derives the existence of the bound (\ref{lambdaExpandingBound}) on the slope of complexity growth.
It is puzzling why this saturation occurs at the equation of state $w = -5/3$. It would be interesting to study this cosmological background further, despite it violating the null energy condition, to see if it has any other cosmological significance.

Let us now consider the behavior of complexity (\ref{ComplexitySqueeze})
for decelerating backgrounds $w > -1/3$.
Recall that in contrast to accelerating backgrounds, modes begin outside the horizon for arbitrarily early times in decelerating backgrounds, and re-enter the horizon at late times.
At early times, then, the squeezing parameter is large $r_k \gg 1$ and growing, as found in (\ref{DecelLargeScaleSolnSqueeze}).
For accelerating backgrounds the squeezing angle asymptotically approaches a constant in the superhorizon limit.
In contrast, for decelerating backgrounds the squeezing drifts away from a constant as the universe expands, as found in (\ref{DecelLargeScaleSolnAngle})
\be
\phi_k(a) \approx -\frac{\pi}{2} + \frac{k\ \eta_0}{|\beta|(|\beta|+2)}\ a^{1/|\beta|}\, .
\label{DecelLargeScaleSolnAngleRepeat}
\ee
The combination of large squeezing and $\phi_k \approx -\pi/2$ again implies that the complexity is large, so that the first term in (\ref{ComplexitySqueeze}) is larger than the second,
and we have, again
\be
{\mathcal C} & \approx & \frac{1}{\sqrt{2}}\,
{\left| \ln \left|
\frac{1+ e^{-2i\phi_k} \,\tanh{r_k}}{1 - e^{-2i\phi_k} \,\tanh{r_k}}
\right| 
\right|}\, .
\label{Capprox2}
\ee
Inserting (\ref{DecelLargeScaleSolnAngleRepeat}) into (\ref{Capprox2}), and using the limit $r_k \gg 1$, the complexity in the early-time, superhorizon limit behaves as
\be
{\mathcal C} \approx -\frac{1+3w}{2\sqrt{2}}\ \ln\left(a/a_*\right)\, ,
\label{DecelComplexity}
\ee
where $a_* \equiv (|\beta|(2+|\beta|)/k\ \eta_0)^{1/|\beta|}$ is a constant, which approximately sets the time of horizon re-entry (notice that when $a\approx a_*$ we have ${\mathcal C} \approx 0$).
The coefficient of (\ref{DecelComplexity}) is negative for decelerating solutions $w > -1/3$, so this implies that the complexity {\bf decreases} as the universe expands as $\sim - \ln a$, in contrast to accelerating backgrounds.
In particular, we can write the slope $\lambda$
\be
\frac{d{\mathcal C}}{dt} \equiv \lambda \approx  -\frac{1+3w}{2\sqrt{2}}\ H\, .
\label{DecelLambda}
\ee
Notice that the slope of the decreasing complexity is {\bf unbounded} from below as the equation of state increases, unlike accelerating backgrounds where the slope of the complexity saturated after the equation of state reached a  threshold value. The slope of the decreasing complexity can be made arbitrarily negative for $w \gg 1$, so that there is no corresponding bound of the form (\ref{lambdaBound}) from below.

\begin{figure}[t]
\includegraphics[width=.7\textwidth]{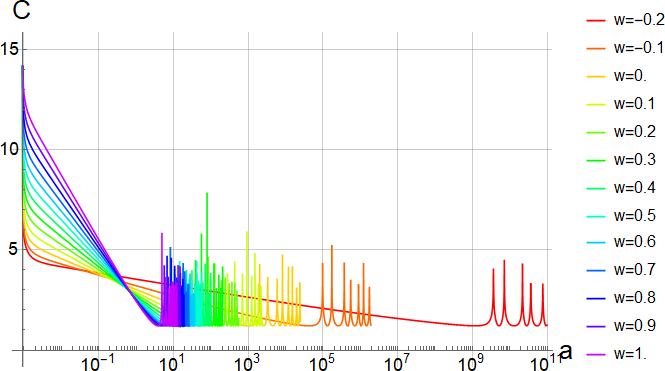}
\caption{The complexity for decelerating expanding solutions decays linearly with $\ln a$ at early times. The slope of the decay depends on the equation of state and is unbounded below, as discussed in Section \ref{sec:Complexity}.}
\label{fig:DecceleratingExpandingComplexity}
\end{figure}

After the mode re-enters the horizon, our analysis from Section \ref{sec:SqueezedCosmo} found that the squeezing parameter ``freezes-in'' at some large value $r_* \gg 1$, while the squeezing angle grows with time as $\phi_k \sim a^{1/|\beta|}$.
Correspondingly, the complexity (\ref{Capprox2}) will oscillate about a fixed ``freeze-in'' value, but will not grow or decay.
Figure \ref{fig:DecceleratingExpandingComplexity} shows the complexity for numerical solutions to the squeezing equations of motion (\ref{eom3})-(\ref{eom4}) for several equations of state corresponding to decelerating backgrounds, illustrating the decreasing complexity and ``freeze-in'' features discussed above, as well as the dependence of the slope on the equation of state.
The panel on the left of Figure \ref{saturation} shows how the negative slope for decelerating backgrounds matches up smoothly to the behavior of $\lambda$ for accelerating backgrounds, and is unbounded below.

Finally, let us briefly comment on the marginal case of $w = -1/3$.
As found in the discussion around (\ref{ExpandingLargeScaleMarginalSqueeze})-(\ref{ExpandingLargeScaleMarginalAngle}), the behavior of these modes depends on whether the mode is inside $k\eta_0 \gg 1$ or outside $k\eta_0 \ll 1$ the horizon.
For modes inside the horizon, the squeezing is small and the squeezing angle is approximately fixed. From (\ref{ComplexitySqueeze}) the complexity is also correspondingly small and approximate constant, up to small oscillations.
For modes outside the the horizon we have the solutions
(\ref{ExpandingLargeScaleMarginalSqueeze})-(\ref{ExpandingLargeScaleMarginalAngle}).
Recall that the squeezing angle is shifted slightly $\phi_k \approx -\frac{\pi}{2} + \frac{1}{2} k\ \eta_0$ in the late-time limit $a \gg 1$.
Because of this slight shift, at late times the complexity approaches a constant ${\mathcal C} \sim \ln \left(|k\ \eta_0|^{-1}\right)/\sqrt{2} \gg 1$.
Correspondingly, the slope at late times vanishes $\lambda \approx 0$, which agrees with the $w\rightarrow -1/3$ limit of (\ref{lambdaExpanding}) and (\ref{DecelLambda}).



%

\subsection{Complexity for Contracting Solutions}

As with expanding solutions, the qualitative nature of the behavior of the complexity (\ref{ComplexitySqueeze}) will depend on whether the contracting universe background is accelerating $w < -1/3$ or deceleration $w > -1/3$.
Let us begin as before with the accelerating contracting backgrounds.
Recall that early times for a contracting background will correspond to large scale factor $a \gg 1$, while at late times the scale factor will be small $a \ll 1$.
At early times, a cosmological mode in an accelerating contracting background begins outside the horizon, with large and growing squeezing parameter $r_k \gg 1$ (\ref{ContractAccelLargeScaleSqueeze}).
The squeezing angle grows slowly on large scales as $\phi_k \approx k\eta_0/(2\beta+1) a^{-1/\beta}$, from
(\ref{ContractAccelLargeScaleAngle}).
Using the approximation (\ref{com1}) 
\begin{eqnarray}
{\mathcal C} \approx \frac{1}{\sqrt{2}}\, 
\left| \ln \left| \frac{1 + e^{-2i\phi_k}\tanh r_k}{1 - e^{2i\phi_k}\tanh r_k}\right| \right|
\end{eqnarray}
for the complexity again,
the complexity for early-time (superhorizon) accelerating contracting backgrounds is
\be
{\mathcal C} \approx
\frac{1}{\sqrt{2}}
\ln\frac{\eta_0}{\eta} = -\frac{1+3w}{2\,{\sqrt{2}}}\, \ln(a/a_*) ~,
\ee
where $a_* = (k\ \eta_0/(2\beta+1))^{1/\beta}$ is again a constant that arises as roughly setting the scale for horizon re-entry.
Since the scale factor is decreasing with time, the complexity is actually decreasing with time.
As in the expanding case, we define the slope $\lambda$ of the complexity
\begin{eqnarray}
\frac{d{\mathcal C}}{dt} \equiv \lambda \approx \frac{1+3w}{2\,{\sqrt{2}}}\,|H| < 0\, ,
\end{eqnarray}
where we have again written $H \equiv \dot a/a$, and used absolute value bars since $H < 0$ for a contracting universe.
Thus the slope is negative for accelerating backgrounds $w < -1/3$, similar to the decelerating expanding backgrounds from the previous subsection.
Notice again that when the complexity is decreasing the slope is unbounded from below, and becomes arbitrarily negative as the equation of state becomes arbitrarily negative. We don't expect an arbitrarily negative equation of state to describe our universe, however, so it is not clear what the physical consequences of this limit are.

\begin{figure}[t]
\includegraphics[width=.7\textwidth]{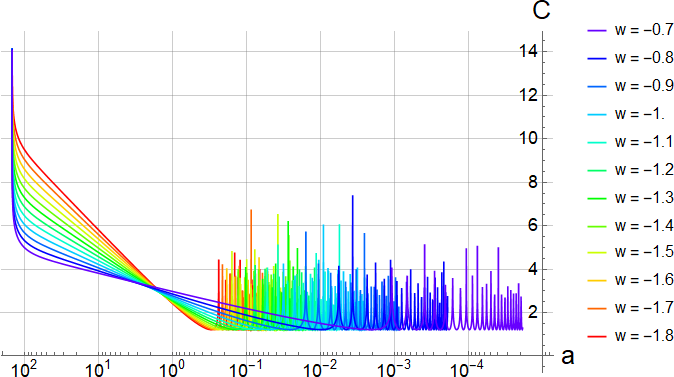}
\caption{The complexity for a accelerating contracting universe ($w < -1/3$) decays with a slope that depends on the equation of state, and is unbounded below, as discussed in Section \ref{sec:Complexity}; when the mode exits the horizon the complexity ``freezes in'', and oscillates about the freeze-in value. Note that since the universe is contracting, the scale factor evolves from large values to small values, thus the flow of time is to the right.}
\label{fig:ContractingAccelComplexity}
\end{figure}

As the modes re-enter the horizon, the squeezing parameter ``freezes-in'' $r_k \approx r_*$ and the squeezing angle begins to run, as found in (\ref{ContractAccelSmallScaleSqueeze})-(\ref{ContractAccelSmallScaleAngle}).
From (\ref{ComplexitySqueeze}), the complexity also ``freezes-in'' and the running squeezing angle causes it to oscillate about the freeze-in value, as seen in the numerical solutions shown in Figure \ref{fig:ContractingAccelComplexity} for several accelerating equations of state.

\begin{figure}[t]
\includegraphics[width=.7\textwidth]{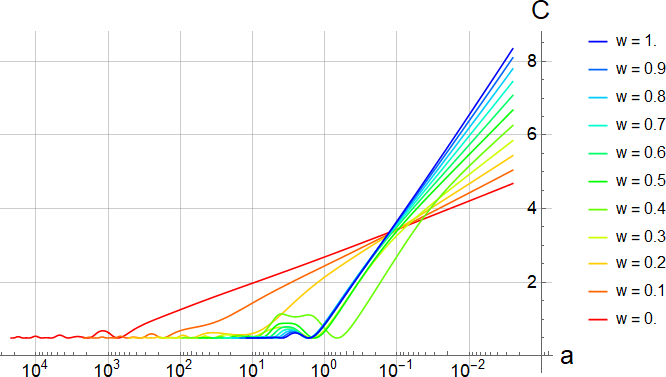}
\caption{The complexity for a decelerating contracting universe ($w > -1/3$) grows with a slope that depends on the equation of state until $w = 1$, after which the slope saturates, as discussed in Section \ref{sec:Complexity}. Note that since the universe is contracting, the scale factor evolves from large values to small values, thus the flow of time is to the right.}
\label{fig:ContractingNonAccelComplexity}
\end{figure}

In contrast to the expanding case, it turns out it is the decelerating $w > -1/3$ contracting backgrounds that will give rise to a growing complexity. 
At early times the squeezing is small $r_k \ll 1$ and the squeezing angle is approximately fixed $\phi_k \approx \frac{\pi}{4}$.
Correspondingly, the complexity (\ref{ComplexitySqueeze}) is also small ${\mathcal C} \ll 1$ while the modes are inside the horizon.
However, at late times $a \ll 1$ as the modes exit the horizon the squeezing parameter begins to grow and the squeezing angle will asymptotically approach zero as found in (\ref{ContractDecelLargeScaleSqueeze})-(\ref{ContractDecelLargeScaleAngle}), so that the complexity (\ref{ComplexitySqueeze}) also begins to grow.
While the growth of the squeezing parameter (\ref{ContractDecelLargeScaleSqueeze}) is universal,
the behavior of the squeezing angle again depends on the equation of state (\ref{ContractDecelLargeScaleAngle}).
Inserting (\ref{ContractDecelLargeScaleSqueeze})-(\ref{ContractDecelLargeScaleAngle}) into (\ref{com1}), the dependence of the complexity on the scale factor grows
as $\sim \ln (1/a)$ at late times
\be
{\mathcal C}
\approx \begin{cases}
\frac{1+3w}{2\,\sqrt{2}}\, 
\ln\left( 1/a\right) + \text{const.} 
~ & -1/3<w<1
\\
\frac{1+3w}{2\sqrt{2}}\, \ln \left(1/a\right)
+ \frac{1}{\sqrt{2}}\,\ln\left(\ln(1/a)\right)~ 
& w=1
\\
\sqrt{2}\,\ln(1/a) + \text{const.}
~ & w>1 
\end{cases}\, ,
\label{ContractComplexity}
\ee
where we wrote the argument of the log as $(1/a)$ to emphasize that as the scale factor $a$ decreases (as the universe contracts) the complexity increases.
As with the accelerating expanding backgrounds, the rate of growth of complexity in the late-time limit $a \ll 1$ depends on the equation of state.
Putting the early- and late-time behaviors together, we see that the complexity for decelerating contracting backgrounds mirrors that of accelerating expanding backgrounds.
At early times, when the mode is inside the horizon the complexity is small; after the mode exits the horizon the complexity grows linearly with $\ln (1/a)$ (as $a \rightarrow 0$), as in (\ref{ContractComplexity}).
Figure \ref{fig:ContractingNonAccelComplexity} shows numerical solutions of the squeezing equations of motion (\ref{eom7})-(\ref{eom8}) inserted into the complexity (\ref{ComplexitySqueeze}), and demonstrates the growth of complexity as modes exit the horizon, and the dependence of the slope of complexity on the equation of state.
As mentioned earlier for expanding solutions, this behavior for the complexity is a characteristic of quantum chaos, suggesting 
that decelerating contracting solutions display some of the essential features of quantum chaos, including a scrambling time and Lyapunov exponent.

While the complexity itself does not saturate, the slope of the growth of complexity -- identified as the Lyapunov exponent in the language of quantum chaos -- does saturate as the equation of state reaches a ``threshold'' value.
In particular, the slope of the growth of the complexity with respect to cosmic time is
\be
\frac{d{\mathcal C}}{dt} \equiv \lambda
\approx \begin{cases}
\frac{1+3w}{2\sqrt{2}}\,|H|~& -1/3<w<1
\\
\frac{1+3w}{2\sqrt{2}}\,|H| 
+ \frac{|H|}{\sqrt{2}\ \ln(a)}
~ & w=1 
\\
\sqrt{2}\ |H| ~ ~ & w>1 
\end{cases}\, .
\label{ContractLambda}
\ee
Again we see that the slope of the growing complexity {\bf saturates}, so that we independently find a bound on the growth of complexity of the form 
\be
\lambda  = \frac{d{\mathcal C}}{dt} \leq \sqrt{2}\ |H|\, .
\label{lambdaContractBound}
\ee
Again, the similarity with the bound \cite{Maldacena:2015waa} (up to an ${\mathcal O}(1)$ factor, as discussed earlier) is remarkable.
Interestingly, the ``threshold'' value of the equation of state for decelerating contracting backgrounds occurs for ``stiff'' matter $w = 1$, with ``ultra-stiff'' matter $w \gg 1$ saturating the bound (\ref{lambdaContractBound}).
This brings to mind the $w \gg 1$ contracting phase of the Ekpyrotic universe \cite{Khoury_2001} (see also the review \cite{Lehners_2008} for more references). Unlike the expanding de Sitter background, then, the $w \gg 1$ contracting background \emph{maximizes} the growth in complexity (\ref{ContractLambda}), thus also maximizing the corresponding Lyapunov exponent in the language of quantum chaos.
It would be interesting to study if this difference in the growth rates of complexity between expanding de Sitter and  contracting ultra-stiff matter backgrounds has any broader implications for the overall generation of complexity for the inflationary and Ekpyrotic scenarios. We leave these interesting questions for future work.

Finally we will briefly note that the marginal equation of state $w=-1/3$ for a contracting universe will behave similarly to the expanding case.
If the mode is initially inside the horizon it will remain inside the horizon, with correspondingly small complexity.
Alternatively, if the mode is outside the horizon it will remain outside the horizon; while the squeezing will grow, the complexity will remain constant because the squeezing angle also becomes constant at late times.

\section{Discussion}
\label{sec:Discussion}

In this paper we have used the language of squeezed states to compute the circuit complexity of scalar cosmological perturbations, as outlined in \cite{us}, for expanding and contracting FLRW backgrounds with fixed equation of state $w$.
We found in Section \ref{sec:SqueezedCosmo} that the behavior of the squeezing parameters follows the general behavior discussed in \cite{us}: when a mode is inside the horizon the squeezing is fixed, up to small oscillations, while the squeezing grows for modes outside the horizon.
We then used this behavior of the squeezing parameters to compute the corresponding cosmological circuit complexity.
The behavior of the complexity depends on whether the cosmological background is expanding or contracting, and whether that background is accelerating $w < -1/3$ or decelerating $w > -1/3$.

The cosmological complexity grows linearly with $\log a$ when modes are outside the horizon for accelerating expanding backgrounds and decelerating contracting backgrounds.
In our model we do not see a saturation in the complexity itself, though this may be due to the expanding background acting as a classical coupling to the effective frequency of the cosmological perturbation.
The linear growth of complexity is a characteristic of a chaotic system, so we propose that cosmological perturbations are chaotic for these corresponding backgrounds.
We independently uncovered a bound on the growth rate of complexity $d{\mathcal C}/dt \leq \sqrt{2}\ |H|$,
which we identify as the Lyapunov exponent in the language of quantum chaos, and is similar up to ${\mathcal O}(1)$ factors to other bounds on the growth of complexity proposed in the literature \cite{Maldacena:2015waa}.
This bound is saturated for equations of state $w < -5/3$ for expanding backgrounds and $w > 1$ for contracting backgrounds.
It is unclear what the independent physical significance of these ``threshold'' equations of state is, and it would be interesting to study this further.
Expanding de Sitter space is ``maximally chaotic'' among backgrounds
that satisfy the null energy condition, in the sense that it has the largest Lyapunov exponent.
Continuing in the language of quantum chaos, we identify the scrambling time as the time that a mode exits the horizon. For de Sitter space, we argue that the maximum scrambling time is $t_{scramb} \sim H_{dS}^{-1}\ \ln(M_{pl}/H_{dS})$ for small wavelength modes, or correspondingly the maximum number of e-folds up to scrambling is $N_e^{scramb} \sim \ln(M_{pl}/H_{dS})$.
These estimates are in line with other estimates of the scrambling time for de Sitter space \cite{Dvali:2013eja,gary}.

For decelerating expanding backgrounds, or accelerating contracting backgrounds, the complexity decays linearly with $\log a$ when the mode is outside the horizon, until eventually ``freezing-in'' to a fixed value up to oscillations when the mode re-enters the horizon.
The slope of the decaying complexity is unbounded from below, so arbitrarily large equations of state ($|w| \gg 1$ with $w > 0$ for decelerating expanding backgrounds or $w < 0$ for accelerating contracting backgrounds) give rise to arbitrarily steep slopes, though it is unclear how physical these backgrounds with extremely large equations of state are.

A number of interesting questions remain. In this and other recent work \cite{us}, we studied cosmological perturbations with sound speed equal to one, corresponding to either perturbations of a canonical scalar field or gravitational wave perturbations.
It would be interesting to study further perturbations with an equation of state that differs from one, such as scalar fields with non-canonical kinetic terms as well as hydrodynamic fluids, such as radiation with $c_s^2 = 1/3$.
We would like to understand better the implications of our observation that for modes outside the horizon the complexity grows for accelerating expanding backgrounds, but decays for decelerating expanding backgrounds (with similar statements for contracting backgrounds). 
This behavior for the complexity could be signaling some amount of entanglement of the perturbations with the degrees of freedom on the horizon.
In particular, an increasing complexity could be signaling increasing entanglement of the mode with the horizon as the mode stretches outside the horizon, while a decreasing complexity signals dis-entanglement as the mode re-enters the horizon. It would be interesting to study this reasoning further.
Finally, we note that in our model the complexity itself is unbounded, and can continue to grow indefinitely for accelerating expanding backgrounds. Perhaps there are other physical limits in our system that bound the complexity and prevent it from becoming arbitrarily large, in line with other expectations on the behavior of complexity.
These and other interesting questions we leave to future work.

\section*{Acknowledgements}
AB is supported by Research Initiation Grant (RIG/0300) provided by IIT-Gandhinagar. This work was supported by the Natural Sciences and Engineering Research Council of Canada. 
\appendix

\section{Intermediate Behavior of Complexity}
\label{app:IntermediateBehavior}

In the investigation of the solutions to the squeezing equations of motion (\ref{eom3})-(\ref{eom4}) for accelerating expanding backgrounds $\beta > 0$ in Section \ref{sec:SqueezedCosmo}, we found that the subleading behavior of the squeezing angle $\phi_k$ on superhorizon scales and late times $a \gg 1$ contained a combination of homogeneous and inhomogeneous solutions
\begin{equation}
\phi_k \approx -\frac{\pi}{2} \ + \ \underbrace{
\frac{k\ |\eta_0|}{2\beta-1}\, \frac{1}{a^{1/\beta}}}_{\text{Inhomogeneous}} +  \underbrace{\frac{B}{a^2}}_{\text{Homogeneous}}, \ \  \beta>1/2\, ,
\label{Intphi}
\end{equation}
where $B$ is a constant set by the initial conditions.
As argued there, for $\beta > 1/2$ the inhomogeneous term dominates at late times $a \gg 1$, while for $\beta < 1/2$ the homogeneous term dominates ($\beta = 1/2$ is a special case, and is considered in more detail in the discussion above (\ref{AccelLargeScaleSolnSqueeze})).
The dominant subleading behavior of the squeezing angle is important because it determines the late-time slope of the growing complexity, as we saw in Section \ref{sec:Complexity}.
The transition to the homogeneous term as $\beta < 1/2$
leads to the saturation of the slope of complexity $d{\mathcal C}/dt \leq \sqrt{2}\ H$.

However, for $\beta > 1/2$ occasionally the combination of the initial conditions set by $B$, and the coefficient on the inhomogeneous term, in (\ref{Intphi}) can be such that the homogeneous term temporarily dominates over the inhomogeneous term in an intermediate regime. This intermediate regime occurs after the mode has exited the horizon, but at still relatively early times.
The subleading behavior of the squeezing angle thus behaves as $\phi_k = -\pi/2 + B/a^2$ even for $\beta > 1/2$, in this intermediate regime.
After a sufficiently long amount of time, however, the inhomogeneous term in (\ref{Intphi}) begins to dominate again, with the transition between the two behaviors occuring at
\begin{equation}
   \frac{k\ |\eta_0|}{2\beta-1}\, \frac{1}{a^{1/\beta}} \sim \frac{B}{a^2}\, .
\end{equation}
After this transition, the squeezing angle is again dominated by the inhomogeneous term at late times for $\beta > 1/2$.

\begin{figure}[t]
\includegraphics[width=.7 \textwidth]{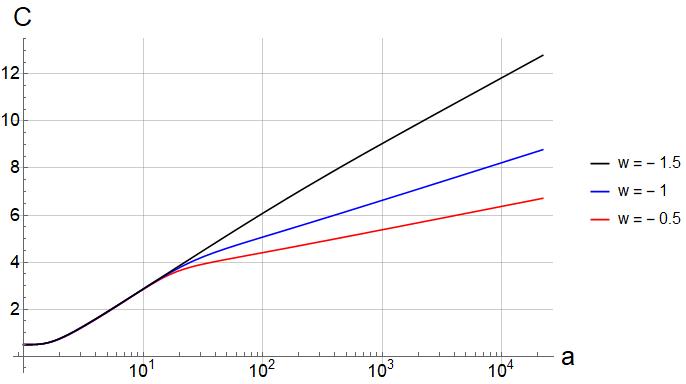}
\caption{For certain initial conditions the subleading behavior of the squeezing angle can temporarily be dominated by the homogeneous term in (\ref{Intphi}) for equations of state $-1/3 < w < -5/3$ in an intermediate regime of the scale factor. The corresponding complexity temporarily saturates the bound (\ref{lambdaExpandingBound}) on the growth of complexity before transitioning to a smaller slope given by (\ref{lambdaExpanding}). This effect is shown here for three sample equations of state, $w = -0.5$, $w=-1$, and $w=-1.5$.}
\label{fig:Intermediate}
\end{figure}

This transition in behavior for the squeezing angle leads to a corresponding change in the slope of the growing complexity.
At early times, when the mode is within the horizon, the complexity is still small. When the mode is in the intermediate regime identified above, however, the slope is now set by the homogeneous term
\be
{\mathcal C} \approx \sqrt{2}\ \ln(a)\, ,
\ee
which saturates the general limit (\ref{lambdaExpandingBound}).
After the transition, however, the slope changes to the (smaller) inhomogeneous value at late times
\be
{\mathcal C} \approx -\frac{1+3w}{2\sqrt{2}}\ \ln(a)\, ,
\ee
(recall that $-1/3 < w < -5/3$ for $\beta > 1/2$, so the coefficient above is positive).
This leads to a ``break'' in the slope of the complexity, as shown in Figure \ref{fig:Intermediate},
which we recognize as simply a transition between the inhomogeneous and homogeneous solutions for the squeezing angle (\ref{Intphi}).
Despite this minor caveat, the late-time slopes identified in (\ref{lambdaExpanding}) are correct since they are taken sufficiently late that only the asymptotic behavior $a \gg 1$ is important.
A similar effect also occurs for the squeezing angle on decelerating contracting backgrounds, but it is identical in nature so we will not consider it in detail further.

\bibliographystyle{utphysmodb}

\bibliography{refs}

\providecommand{\href}[2]{#2}\begingroup\raggedright\begin{thebibliography}{10}

\bibitem{com1}
L.~Susskind,  {\em {Computational Complexity and Black Hole Horizons}},
  Fortsch. Phys. {\bf 64} (2016) 24--43
  [\href{http://www.arXiv.org/abs/1403.5695}{{\tt 1403.5695}}], [Addendum:
  Fortsch.Phys. 64, 44--48 (2016)].

\bibitem{com2}
D.~Stanford and L.~Susskind,  {\em {Complexity and Shock Wave Geometries}},
  Phys. Rev. D {\bf 90} (2014), no.~12, 126007
  [\href{http://www.arXiv.org/abs/1406.2678}{{\tt 1406.2678}}].

\bibitem{com3}
A.~R. Brown, D.~A. Roberts, L.~Susskind, B.~Swingle and Y.~Zhao,  {\em
  {Holographic Complexity Equals Bulk Action?}}, Phys. Rev. Lett. {\bf 116}
  (2016), no.~19, 191301 [\href{http://www.arXiv.org/abs/1509.07876}{{\tt
  1509.07876}}].

\bibitem{com4}
A.~R. Brown, D.~A. Roberts, L.~Susskind, B.~Swingle and Y.~Zhao,  {\em
  {Complexity, action, and black holes}}, Phys. Rev. D {\bf 93} (2016), no.~8,
  086006 [\href{http://www.arXiv.org/abs/1512.04993}{{\tt 1512.04993}}].

\bibitem{MyersCC}
R.~Jefferson and R.~C. Myers,  {\em {Circuit complexity in quantum field
  theory}}, JHEP {\bf 10} (2017) 107
[\href{http://www.arXiv.org/abs/1707.08570}{{\tt 1707.08570}}].

\bibitem{Guo:2018kzl}
M.~Guo, J.~Hernandez, R.~C. Myers and S.-M. Ruan,  {\em {Circuit Complexity for
  Coherent States}}, JHEP {\bf 10} (2018) 011
[\href{http://www.arXiv.org/abs/1807.07677}{{\tt 1807.07677}}].

\bibitem{Khan:2018rzm}
R.~Khan, C.~Krishnan and S.~Sharma,  {\em {Circuit Complexity in Fermionic
  Field Theory}}, Phys. Rev. D {\bf 98} (2018), no.~12, 126001
  [\href{http://www.arXiv.org/abs/1801.07620}{{\tt 1801.07620}}].

\bibitem{Hackl:2018ptj}
L.~Hackl and R.~C. Myers,  {\em {Circuit complexity for free fermions}}, JHEP
  {\bf 07} (2018) 139 [\href{http://www.arXiv.org/abs/1803.10638}{{\tt
  1803.10638}}].

\bibitem{Bhattacharyya:2018bbv}
A.~Bhattacharyya, A.~Shekar and A.~Sinha,  {\em {Circuit complexity in
  interacting QFTs and RG flows}}, JHEP {\bf 10} (2018) 140
[\href{http://www.arXiv.org/abs/1808.03105}{{\tt 1808.03105}}].

\bibitem{Camargo:2018eof}
H.~A. Camargo, P.~Caputa, D.~Das, M.~P. Heller and R.~Jefferson,  {\em
  {Complexity as a novel probe of quantum quenches: universal scalings and
  purifications}}, Phys. Rev. Lett. {\bf 122} (2019), no.~8, 081601
  [\href{http://www.arXiv.org/abs/1807.07075}{{\tt 1807.07075}}].

\bibitem{Ali:2018aon}
T.~Ali, A.~Bhattacharyya, S.~Shajidul~Haque, E.~H. Kim and N.~Moynihan,  {\em
  {Post-Quench Evolution of Distance and Uncertainty in a Topological System:
  Complexity, Entanglement and Revivals}},
\href{http://www.arXiv.org/abs/1811.05985}{{\tt 1811.05985}}.

\bibitem{NL3}
M.~R. Nielsen, M.~A.and~Dowling,  {\em {The geometry of quantum computation}},
  Science {\bf 311} (2006), no.~4, 1133--1135
[\href{http://www.arXiv.org/abs/0701004}{{\tt 0701004}}].

\bibitem{Chapman:2017rqy}
S.~Chapman, M.~P. Heller, H.~Marrochio and F.~Pastawski,  {\em {Toward a
  Definition of Complexity for Quantum Field Theory States}}, Phys. Rev. Lett.
  {\bf 120} (2018), no.~12, 121602
  [\href{http://www.arXiv.org/abs/1707.08582}{{\tt 1707.08582}}].

\bibitem{Caputa:2017yrh}
P.~Caputa, N.~Kundu, M.~Miyaji, T.~Takayanagi and K.~Watanabe,  {\em {Liouville
  Action as Path-Integral Complexity: From Continuous Tensor Networks to
  AdS/CFT}}, JHEP {\bf 11} (2017) 097
  [\href{http://www.arXiv.org/abs/1706.07056}{{\tt 1706.07056}}].

\bibitem{Bhattacharyya:2018wym}
A.~Bhattacharyya, P.~Caputa, S.~R. Das, N.~Kundu, M.~Miyaji and T.~Takayanagi,
  {\em {Path-Integral Complexity for Perturbed CFTs}}, JHEP {\bf 07} (2018) 086
  [\href{http://www.arXiv.org/abs/1804.01999}{{\tt 1804.01999}}].

\bibitem{Caputa:2018kdj}
P.~Caputa and J.~M. Magan,  {\em {Quantum Computation as Gravity}}, Phys. Rev.
  Lett. {\bf 122} (2019), no.~23, 231302
  [\href{http://www.arXiv.org/abs/1807.04422}{{\tt 1807.04422}}].

\bibitem{Bhattacharyya:2019kvj}
A.~Bhattacharyya, P.~Nandy and A.~Sinha,  {\em {Renormalized Circuit
  Complexity}}, Phys. Rev. Lett. {\bf 124} (2020), no.~10, 101602
[\href{http://www.arXiv.org/abs/1907.08223}{{\tt 1907.08223}}].

\bibitem{Caputa:2020mgb}
P.~Caputa and I.~MacCormack,  {\em {Geometry and Complexity of Path Integrals
  in Inhomogeneous CFTs}}, \href{http://www.arXiv.org/abs/2004.04698}{{\tt
  2004.04698}}.

\bibitem{Flory:2020eot}
M.~Flory and M.~P. Heller,  {\em {Complexity and Conformal Field Theory}},
  \href{http://www.arXiv.org/abs/2005.02415}{{\tt 2005.02415}}.

\bibitem{Erdmenger:2020sup}
J.~Erdmenger, M.~Gerbershagen and A.-L. Weigel,  {\em {Complexity measures from
  geometric actions on Virasoro and Kac-Moody orbits}},
  \href{http://www.arXiv.org/abs/2004.03619}{{\tt 2004.03619}}.

\bibitem{Balasubramanian:2019wgd}
V.~Balasubramanian, M.~Decross, A.~Kar and O.~Parrikar,  {\em {Quantum
  Complexity of Time Evolution with Chaotic Hamiltonians}}, JHEP {\bf 01}
  (2020) 134 [\href{http://www.arXiv.org/abs/1905.05765}{{\tt 1905.05765}}].

\bibitem{me3}
T.~Ali, A.~Bhattacharyya, S.~S. Haque, E.~H. Kim, N.~Moynihan and J.~Murugan,
  {\em {Chaos and Complexity in Quantum Mechanics}}, Phys. Rev. {\bf D101}
  (2020), no.~2, 026021
[\href{http://www.arXiv.org/abs/1905.13534}{{\tt 1905.13534}}].

\bibitem{Yang:2019iav}
R.-Q. Yang and K.-Y. Kim,  {\em {Time evolution of the complexity in chaotic
  systems: concrete examples}}, JHEP {\bf 05} (2020) 045
  [\href{http://www.arXiv.org/abs/1906.02052}{{\tt 1906.02052}}].

\bibitem{Grishchuk}
L.~P. Grishchuk and Y.~V. Sidorov,  {\em Squeezed quantum states of relic
  gravitons and primordial density fluctuations}, Phys. Rev. D {\bf 42} (Nov,
  1990) 3413--3421.

\bibitem{Albrecht}
A.~Albrecht, P.~Ferreira, M.~Joyce and T.~Prokopec,  {\em {Inflation and
  squeezed quantum states}}, Phys. Rev. {\bf D50} (1994) 4807--4820
[\href{http://www.arXiv.org/abs/astro-ph/9303001}{{\tt astro-ph/9303001}}].

\bibitem{Martin1}
J.~Martin,  {\em {Inflationary perturbations: The Cosmological Schwinger
  effect}}, Lect. Notes Phys. {\bf 738} (2008) 193--241
[\href{http://www.arXiv.org/abs/0704.3540}{{\tt 0704.3540}}].

\bibitem{Martin2}
J.~Martin,  {\em {Cosmic Inflation, Quantum Information and the Pioneering Role
  of John S Bell in Cosmology}}, Universe {\bf 5} (2019), no.~4, 92
[\href{http://www.arXiv.org/abs/1904.00083}{{\tt 1904.00083}}].

\bibitem{us}
A.~Bhattacharyya, S.~Das, S.~Shajidul~Haque and B.~Underwood,  {\em
  {Cosmological Complexity}},
\href{http://www.arXiv.org/abs/2001.08664}{{\tt 2001.08664}}.

\bibitem{Maldacena:2015waa}
J.~Maldacena, S.~H. Shenker and D.~Stanford,  {\em {A bound on chaos}}, JHEP
  {\bf 08} (2016) 106 [\href{http://www.arXiv.org/abs/1503.01409}{{\tt
  1503.01409}}].

\bibitem{Dvali:2013eja}
G.~Dvali and C.~Gomez,  {\em {Quantum Compositeness of Gravity: Black Holes,
  AdS and Inflation}}, JCAP {\bf 01} (2014) 023
  [\href{http://www.arXiv.org/abs/1312.4795}{{\tt 1312.4795}}].

\bibitem{gary}
L.~Aalsma and G.~Shiu,  {\em {Chaos and complementarity in de Sitter space}},
\href{http://www.arXiv.org/abs/2002.01326}{{\tt 2002.01326}}.

\bibitem{Mukhanov}
V.~F. Mukhanov, H.~Feldman and R.~H. Brandenberger,  {\em {Theory of
  cosmological perturbations. Part 1. Classical perturbations. Part 2. Quantum
  theory of perturbations. Part 3. Extensions}}, Phys. Rept. {\bf 215} (1992)
  203--333.

\bibitem{me1}
T.~Ali, A.~Bhattacharyya, S.~Shajidul~Haque, E.~H. Kim and N.~Moynihan,  {\em
  {Time Evolution of Complexity: A Critique of Three Methods}}, JHEP {\bf 04}
  (2019) 087
[\href{http://www.arXiv.org/abs/1810.02734}{{\tt 1810.02734}}].

\bibitem{Bedroya:2019snp}
A.~Bedroya and C.~Vafa,  {\em {Trans-Planckian Censorship and the Swampland}},
  \href{http://www.arXiv.org/abs/1909.11063}{{\tt 1909.11063}}.

\bibitem{Bedroya:2019tba}
A.~Bedroya, R.~Brandenberger, M.~Loverde and C.~Vafa,  {\em {Trans-Planckian
  Censorship and Inflationary Cosmology}}, Phys. Rev. D {\bf 101} (2020),
  no.~10, 103502 [\href{http://www.arXiv.org/abs/1909.11106}{{\tt
  1909.11106}}].

\bibitem{Gibbons:1977mu}
G.~Gibbons and S.~Hawking,  {\em {Cosmological Event Horizons, Thermodynamics,
  and Particle Creation}}, Phys. Rev. D {\bf 15} (1977) 2738--2751.

\bibitem{Khoury_2001}
J.~Khoury, B.~A. Ovrut, P.~J. Steinhardt and N.~Turok,  {\em Ekpyrotic
  universe: Colliding branes and the origin of the hot big bang}, Physical
  Review D {\bf 64} (Nov, 2001).

\bibitem{Lehners_2008}
J.-L. Lehners,  {\em Ekpyrotic and cyclic cosmology}, Physics Reports {\bf 465}
  (Sep, 2008) 223–263.

\end{thebibliography}\endgroup

\end{document}